\documentclass[notitlepage,nofootinbib]{revtex4-1}
\usepackage[latin9]{inputenc}
\usepackage{amsmath}
\usepackage{graphicx}
\usepackage{comment}
\usepackage{xcolor}
\usepackage{hyperref}
\usepackage{diagbox}
\usepackage[normalem]{ulem}

\begin{document}

\title{The prior dependence of the DESI results}

\author{Vrund Patel$^{1}$, Amlan Chakraborty$^{2}$, Luca~Amendola$^{1}$}
\affiliation{$^{1}$Institut f\"ur Theoretische Physik, Universit\"at Heidelberg,
Philosophenweg 16, 69120 Heidelberg, Germany\\ $^{2}$ Indian Institute of Astrophysics, Bengaluru, Karnataka 560034, India
}

\begin{abstract}
    In this short contribution, we explore the prior dependence of the recent DESI results, according to which a Bayesian comparison of $\Lambda$CDM with $w_0w_a$CDM cosmologies favors dynamical dark energy.
   It is a simple exercise to show that extending the prior lower bound  of the  dark energy parameters tilts the preference in favor of $\Lambda$CDM. Adopting the PantheonPlus supernovae catalog, for instance, a shift of the lower bound for $w_0,w_a$ from $-3$ to beyond $-4.6$ and $-5$, respectively, is sufficient to favor $\Lambda$CDM. This calls for caution when interpreting DESI results in the Bayesian context.
\end{abstract}
\maketitle

\section{Introduction}

The recent DESI paper \cite{DESI:2024mwx} found the remarkable result that $\Lambda$CDM is disfavored with respect to a cosmology with dark energy equation of state parameterized as $w(z)=w_0+w_a z/(1+z)$  \cite{2003PhRvL..90i1301L, 2001IJMPD..10..213C}.
A model comparison with the Bayes' ratio finds that the evidence in favor of $w_0w_a$CDM ranges from ``weak" to ``moderate" in Jeffreys' scale language. This result has already generated a heated discussion concerning theoretical and data-analysis aspects.

It is well known that Bayes' ratio depends on the chosen prior. In many fields of research, this might not be a serious problem since priors are often based on sound physical or logical grounds. Cosmology, however, often deals with models that can hardly be tested locally and, therefore, receive little guidance from established physics. Several popular parametrizations of the dark energy equation of state, like  $w_0,w_a$,  are primary examples of this, being introduced on a purely phenomenological basis. For instance, Planck18 (P18) \cite{Planck:2018vyg} and the Dark Energy Survey (DES) \cite{DES:2024tys} adopt different priors for $w_{0}$ and $w_{a}$. P18 uses flat priors of $[-3, 1]$ for $w_{0}$ and $[-5, 5]$ for $w_{a}$, while DES uses $[-10, 5]$ and $[-20, 10]$ for $w_{0}$ and $w_{a}$, respectively. As a consequence, the $w_0,w_a$ priors are largely arbitrary, at least at the lower limit.

In this short contribution, we put on a quantitative basis the dependence of DESI results on the prior, already briefly mentioned in their paper. 
We find that extending the lower bound on $w_0,w_a$ from $-3$ to $-4.6$ and $-5$, respectively, tilts the evidence of DESI+CMB+PantheonPlus in favor of $\Lambda$CDM. The other SN datasets require larger but not unreasonable values to favor $\Lambda$CDM.

This extremely simple exercise is meant mostly to point out the issue in the most transparent way, rather than to find solutions. 

\section{Bayes' ratio for uniform prior}

Bayes' ratio (or Bayes' factor) is a widely used statistical tool for model comparison in cosmology. It is defined as the ratio of the evidence for the models under consideration. Denoting the competing models with subscripts 1 and 2, we have
\begin{equation}
    B_{21} = \frac{\mathcal{E}_{2}}{\mathcal{E}_{1}} = \frac{\int \mathcal{L}_{2} \mathcal{P}_{2}}{\int \mathcal{L}_{1} \mathcal{P}_{1}}
\end{equation}
The integrals extend over the entire domain of the model parameters. Here, $\mathcal{E} = \int \mathcal{LP}$ describes the evidence for the model, where $\mathcal{L}$ is the likelihood derived from the data and the assumed underlying theoretical model, and $\mathcal{P}$ is the prior, denoting the information about the parameters before the experiment. 
Bayes' factor $B_{21}$ is then usually interpreted using Jeffreys' scale \cite{Jeffreys1939-JEFTOP-5}.

If uniform priors $\mathcal{U} = \frac{1}{b - a} = \frac{1}{d}$ are used for the parameters, the expression for the Bayes' factor simplifies to:
\begin{equation}
    B_{21} = \frac{\mathcal{P}_{2} \int \mathcal{L}_{2}}{\mathcal{P}_{1} \int \mathcal{L}_{1}} = \frac{\mathcal{P}_{2}}{\mathcal{P}_{1}} \Gamma = \frac{(d_{\nu_{1}}d_{\nu_{2}}d_{\nu_{3}}...)_{1}}{(d_{\nu_{1}}d_{\nu_{2}}d_{\nu_{3}}...)_{2}} \Gamma
\end{equation}
Here, $\Gamma = \frac{\int \mathcal{L}_{2}}{\int \mathcal{L}_{1}}$, and $\nu_{i}$ represents the $i$-th parameter of the model.  Here we are implicitly assuming that the likelihood integral converges and does not depend any longer on the prior. This will be numerically verified below.

Often, especially in cosmology, the prior range is not built on solid and univocal physical grounds. This is, for instance, the case for the dark energy equation of state. Then,
one might choose to define a new extended range for the uniform prior as $d^{*} = (b - a) + |n| = d + |n|$ and setting $|n| = (\alpha - 1) (b - a)$ (such that $d^{*} \geq d$ and $\alpha \geq 1$), we get for each parameter $d^{*} = \alpha (b-a) = \alpha d$. Thus, the relationship between the new $B^{*}_{21}$ and the $\alpha$'s is given by:
\begin{equation}\label{Eq:new_beta} 
    \log{B^{*}_{21}} = \log \left[{\frac{(\alpha_{\nu_{1}} \alpha_{\nu_{2}} \alpha_{\nu_{3}}...)_{1} (d_{\nu_{1}}d_{\nu_{2}}d_{\nu_{3}}...)_{1}}{(\alpha_{\nu_{1}} \alpha_{\nu_{2}} \alpha_{\nu_{3}}...)_{2} (d_{\nu_{1}}d_{\nu_{2}}d_{\nu_{3}}...)_{2}} \Gamma} \right] = \log \left[ \frac{(\alpha_{\nu_{1}} \alpha_{\nu_{2}} \alpha_{\nu_{3}}...)_{1}}{(\alpha_{\nu_{1}} \alpha_{\nu_{2}} \alpha_{\nu_{3}}...)_{2}} \right] + \log{B_{21}}
\end{equation}

\section{Bayes' ratio in DESI}

The DESI paper \cite{DESI:2024mwx} considered several cosmological models. The comparison of  $\Lambda$CDM (model 1) with  $w_{0}w_{a}$CDM (model 2) produced some eye-catching results. The two models share common parameters, the Hubble constant ($H_{0}$) and current matter density ($\Omega_{m}$), while $w_{0}w_{a}$CDM includes two additional parameters for a time-varying equation of state, $w_{0}$ and $w_{a}$. The CMB parameters included in the DESI analysis are also common to both models.

Ref. \cite{DESI:2024mwx} adopted the priors outlined in Tab. \ref{tab:prior} to compare the two models. They obtained the results shown in Tab. \ref{tab:bayes}, according to which the combination of DESI+CMB data with Supernovae(SN) data from PantheonPlus provides weak evidence favoring $w_{0}w_{a}$CDM over $\Lambda$CDM. When the SN catalog is Union3 or DESY5, the data gives moderate evidence for $w_{0}w_{a}$CDM.

\begin{table}
    \centering
    \begin{minipage}{.48\textwidth}
        \centering
        \begin{tabular}{c||c}
        \hline
            Parameter & Prior \\
            \hline
            $H_{0}$ & $\mathcal{U}[20, 100]$ \\
            $\Omega_{m}$ & $\mathcal{U}[0.01, 0.99]$ \\
            $w_{0}$ & $\mathcal{U}[-3, 1]$ \\
            $w_{a}$ & $\mathcal{U}[-3, 2]$ \\
            \hline
        \end{tabular}
        \caption{\label{tab:prior} Uniform priors used for various parameters in \cite{DESI:2024mwx}.}
    \end{minipage}
    \hfill
     \begin{minipage}{.48\textwidth}
        \centering
        \begin{tabular}{c||c|c}
        \hline
            $| \log B_{21} |$ & SN dataset & Strength for $w_{0}w_{a}$CDM \\
            \hline
            0.65 & PantheonPlus & Weak\\
            2.4 & Union3 & Moderate\\
            2.8 & DESY5 & Moderate\\
            \hline
        \end{tabular}
        \caption{\label{tab:bayes} Values of $|\log B_{21}|$ and their interpretation according to Jeffreys' scale \cite{Trotta:2005ar} from the DESI paper. The second column lists the SN data combined with DESI+CMB.}
    \end{minipage}%
   
\end{table}

Several attempts have been made to verify whether the surprising results are artefacts of the parametrization or of the statistical methods used for analysis, see e.g.  \cite{2024arXiv240504216C,2024arXiv240513588L}. Ref. \cite{2024arXiv240500502P} presented results similar to DESI but using a different dataset consisting of P18+lensing+non-CMB data. The non-CMB data includes Pantheon+ SNIa and BAO data, excluding the recent DESI BAO measurements.  However, we believe the critical question is the justification of the priors (see, for instance, the discussion in \cite{Cortes:2024lgw}) and how results might change when different priors are employed. Since the DESI analysis uses uniform priors, $\log B_{21}$ changes as per Eq. \ref{Eq:new_beta} when updated priors are used. Using this equation, the $\alpha$ factors for the common parameters cancel out, leaving only those for $w_{0}$ and $w_{a}$. We can, therefore, immediately estimate the new Bayes' ratio when the priors become broader. We only extend them from the lower end (i.e., $n<0$) to keep the past evolution of the universe essentially unchanged. 

\begin{figure}
\begin{centering}
\includegraphics[width=15cm]{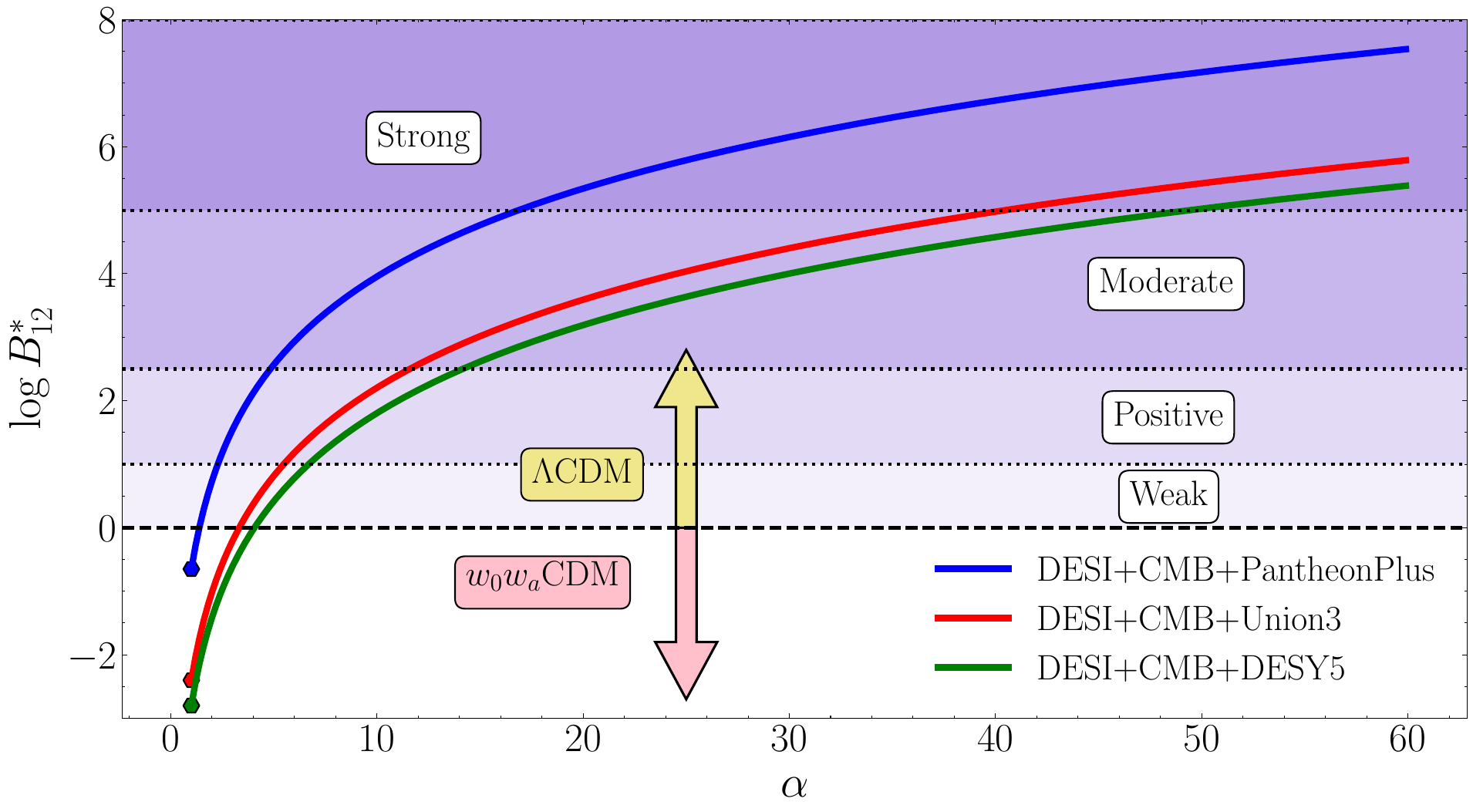}
\par\end{centering}
\caption{\label{fig:prior_all} Distribution of $\log B^{*}_{12}$ for $\alpha$ in the range $[1, 60]$. The colored regions indicate the strength of evidence for $\Lambda$CDM against $w_{0}w_{a}$CDM using Jeffreys' scale. The filled hexagon points at the start of the curves represent the original values from the DESI paper. 
}
\end{figure}

Choosing for simplicity a single broadening factor, we put $\alpha_{w_{0}} = \alpha_{w_{a}} = \alpha$, so that Eq. \ref{Eq:new_beta} simplifies to:
\begin{equation}\label{Eq:final_beta}
    \log{B^{*}_{21}} =   \log{B_{21}} - 2 \log \alpha
\end{equation}
Fig. \ref{fig:prior_all} illustrates the behavior of $\log B^{*}_{12}$ as $\alpha$ increases. A positive  $\log B^{*}_{12}$  (opposite of $\log B^{*}_{21}$) indicates that the data favor $\Lambda$CDM against $w_{0}w_{a}$CDM. An arbitrarily large factor $\alpha$ will always favor the model with fewer parameters.

From Fig. \ref{fig:prior_all} we see that varying the priors in the DESI analysis could switch the preferred model from $w_0w_a$CDM to $\Lambda$CDM. If the DESI+CMB+PantheonPlus analysis is repeated with the lower values of $w_{0}$ and $w_{a}$ extended from -3 to -4.6 and -5, respectively, the data will start favoring $\Lambda$CDM, thereby challenging the current results.
For $\alpha > 14.2$, all SN data support $\Lambda$CDM with moderate strength, and with $\alpha > 16.9$, DESI+CMB with PantheonPlus would strongly favor $\Lambda$CDM over $w_{0}w_{a}$CDM. Tab. \ref{tab:n_values} shows the critical values of the lower bound of the updated priors $a^{*}=a+n$. 

\begin{table}[]
    \centering
    \begin{tabular}{c||c|c|c}
        \hline
             \diagbox{SN data}{$w_{0}/w_{a}$} & $ a^{*}_{\text{Turnaround}}$  & $a^{*}_{\text{Moderate}}$ & $a^{*}_{\text{Strong}}$  \\
            \hline
            PantheonPlus & -4.6/-5 & -18.6/-22.5 & -66.6/-82.5 \\
            Union3 & -12.6/-15 & -45.4/-56 & -161/-200.5 \\
            DESY5 & -15.4/-18.5 & -55.8/-69 & -197/-245.5 \\
            \hline
        \end{tabular}
        \caption{\label{tab:n_values} 
        The critical values of the prior lower bound $a^{*} = a + n$ at which $\log B^{*}_{12}$ values transition into different regions of Jeffreys' scale are shown here for both parameters $w_{0}$ and $w_{a}$.}
\end{table}

\begin{figure}
\begin{centering}
\includegraphics[width=15cm]{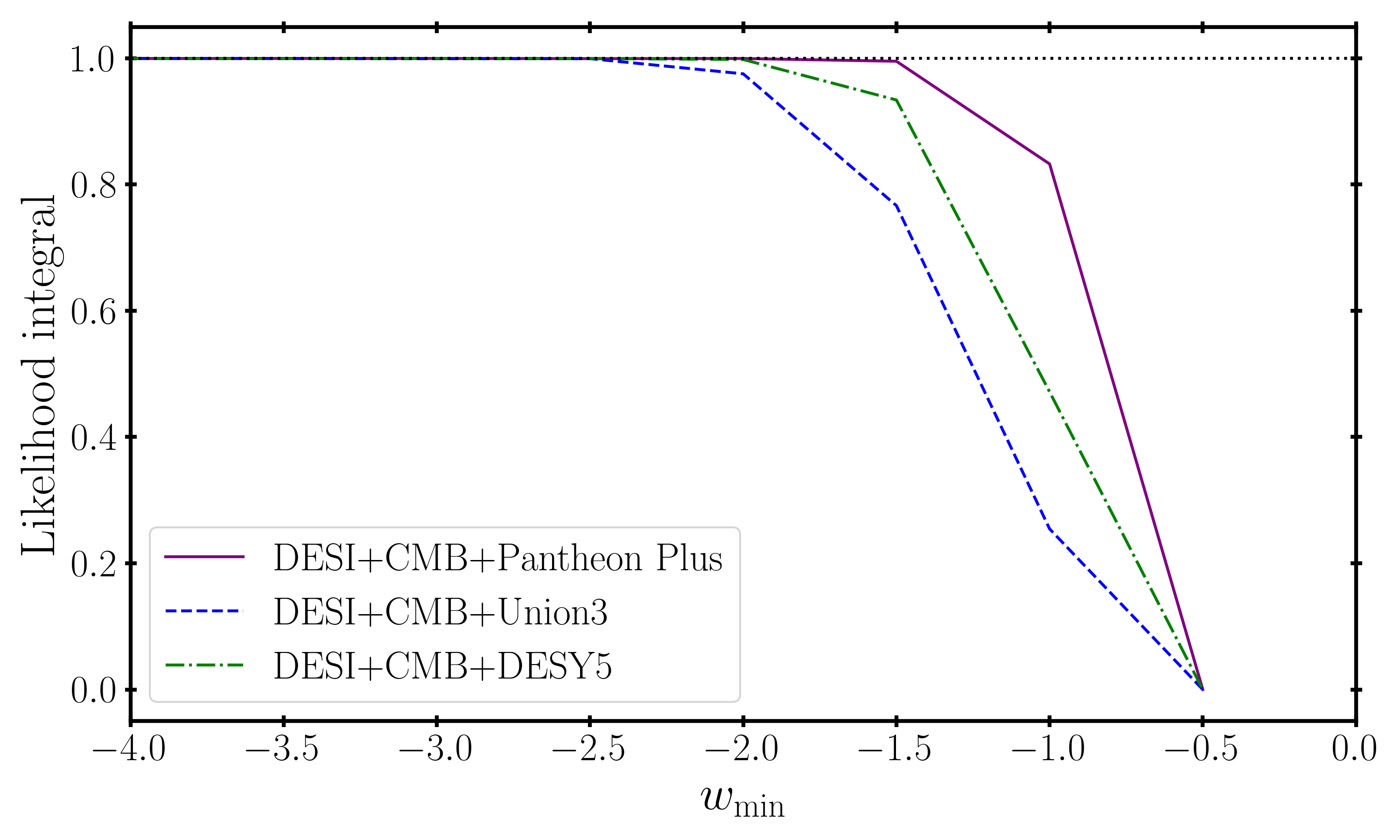}
\par\end{centering}
\caption{\label{fig:likelihood_integral} Integration of the likelihood (arbitrarily normalized to unity when $w_{\rm min}=-4$) is shown for varying values of $w_{\rm min}$, being the lower bound of both $w_0$ and $w_a$  using the same combination of data sets of the DESI paper. The integral shows convergence for $w_{\rm{min}}<-2.5$ for all the data set. 
}
\end{figure}

We further validated our analysis by conducting Markov Chain Monte Carlo (MCMC) simulations utilizing the $w_{0}w_{a}$CDM model, using the MCMC sampler Cobaya \cite{Torrado:2020dgo} and incorporating a composite dataset that includes DESI+CMB, Pantheon Plus, Union3, and DESY5. This approach mirrors the methodology employed by the DESI collaboration \cite{DESI:2024mwx} but with a notable modification: the extension of the parameter space boundaries for $w_0$ and $w_a$ to $-5$, keeping the upper limit same as DESI results, i.e. $1$ for $w_0$ and $2$ for $w_a$. Following the acquisition of results, we proceeded to calculate the integration of the likelihood function, as depicted in Figure \ref{fig:likelihood_integral}. This step is imperative for the computation of the Bayesian evidence. The integration was executed by adjusting the lower limit of the parameters $w_0$ and $w_a$ within the range of $-0.5$ to $-4.0$, as represented by $w_{\rm{min}}$ in Fig. \ref{fig:likelihood_integral} while keeping the upper limits unchanged. The convergence of the integral for $w_{\rm{min}}<-2.5$ across all dataset combinations, as illustrated in the figure, confirms that the determination of evidence for any expanded prior range is contingent upon the parameter $\alpha$.

\section{Discussion}

This simple exercise is not meant to claim that DESI data are actually in favor of $\Lambda$CDM, but merely to point out the consequence of lacking a proper physical justification for the adopted prior. Unfortunately, this issue is hardly solvable within cosmology itself, since many models are entirely phenomenological or very speculative, rather than built on solid physical grounds.

In this short paper, we do not wish to present firm conclusions. We only refer the reader to our paper \cite{2024arXiv240400744A} in which we discuss a radical solution to the prior problem in the Bayesian context: abandon Bayesian model comparison and resort to frequentist hypothesis testing. 

\section*{Acknowledgments}

LA acknowledges support by the Deutsche Forschungsgemeinschaft (DFG, German Research Foundation) under Germany's Excellence Strategy EXC 2181/1 - 390900948 (the Heidelberg STRUCTURES Excellence Cluster). We thank Raul Abramo, Adri\'{a} Gomez-Valent, Savvas Nesseris, Bj\"{o}rn M. Sch\"{a}fer, Arman Shafieloo, and Do\u{g}a Veske for useful comments. AC acknowledges the support by DST-DAAD Indo-German joint research collaboration grant DST/INT/DAAD/P-07/2023(G).

\bibliographystyle{hunsrt}
\bibliography{references}

\end{document}